\documentclass[aps,prd,twocolumn,showpacs]{revtex4}
\usepackage{graphics}

\newcommand{\eref}[1]{Eq. (\ref{#1})}
\newcommand{\fref}[1]{Fig. (\ref{#1})}
\newcommand{\mycite}[1]{Ref. \cite{#1}}
\newcommand{\beq}{\begin{equation}}
\newcommand{\eeq}{\end{equation}}
\newcommand{\bea}{\begin{eqnarray}}
\newcommand{\eea}{\end{eqnarray}}
\newcommand{\sstw}{\ensuremath{{\rm sin}^2\theta_W}}

\begin{document}

\title{Neutron Form Factor from Neutrino-Nucleus Coherent Elastic Scattering}

\author{Philip S. Amanik and Gail C. McLaughlin}
\affiliation{Department of Physics, North Carolina State University, Raleigh, NC 27695}

\date{\today}

\begin{abstract}
We analyze the prospect of measuring the neutron form factor of a nucleus through the detection of neutrino-nucleus coherent elastic scattering. We predict numbers of events in a liquid noble nuclear recoil detector at a stopped pion neutrino source. We discuss the precision required to distinguish between different theoretical models for the form factor.
\end{abstract} 


\maketitle

\section{Introduction}
Hadronic (pion, proton) scattering experiments have given us our first glimpse of the neutron distributions within nuclei. 
However, due to uncertainties and disagreements in theoretical fit calculations, the neutron radii of medium to heavy nuclei are still not sufficiently well known\cite{HPSM,furnstahl}. Proton charge radii are accurately known from electron scattering\cite{HPSM}, but until we have a better understanding of neutron radii our picture of the nucleus will be incomplete. The neutron(proton) radius is related to the neutron(proton) density distribution in a nucleus.  An accurate measurement of the neutron density distribution is desirable not only for the sake of having a more complete picture of the nucleus, such knowledge will be applicable in other areas of physics as well. A precise measurement of the density of a nucleus will provide a better understanding of the saturation density of nuclear matter. Hence, a better knowledge of neutron densities could be useful for applications ranging from reducing uncertainties in atomic parity violation experiments to understanding properties of neutron stars (see \cite{HPSM,furnstahl} and references therein.) 

In the expression for the cross section for a projectile scattering with a nucleus, the form factor gives a measure of the charge distribution in the nucleus. Specifically, the form factor is the Fourier transform of the density distribution of neutrons and protons.  A measurement of the neutron form factor of the lead nucleus ($^{208}{\rm Pb}$) using parity violating electron scattering has been discussed\cite{HPSM} and an experiment tentatively scheduled at JLAB for 2008\cite{prex}.  

In this paper, we present another way the neutron form factor of a nucleus could be measured --- using neutrino-nucleus elastic scattering.  The matrix element for the process of neutrino-nucleus elastic scattering\cite{freedman} is obtained by the superposition principle \cite{FST}: the individual amplitudes (with relative phase factors) for the neutrino to scatter off each nucleon are added and then the total is squared.  For the case of a spin zero nucleus, ignoring radiative corrections, the differential cross section is given by \cite{BMR}
\beq
\frac{d\sigma}{dT} = \frac{G_F^2}{2\pi} M \left [ 
2 - \frac{2T} E + \left(\frac T E\right)^2 - \frac{M T} {E^2} \right ] \frac{Q_W^2}{4} F^2(Q^2). 
\label{eq:cross-section}
\eeq
In this expression, $G_F$ is the Fermi constant, $E$ is the neutrino energy, $T$ is the nuclear recoil energy,
$M$ is the mass of the nucleus, $Q_W = N- Z(1-4\sstw)$ is the weak charge ($\sstw\approx0.231$), 
$Q^2 = 2E^2TM/(E^2 - ET)$ is the squared momentum transfer and $F(Q^2)$ is the form factor.  
Denoting the neutron and proton densities as $\rho_{n,p}(r)$, the form factor is \cite{HCM}
\beq
F(Q^2) = \frac 1 {Q_W}\int [\rho_n(r) - (1-4\sstw)\rho_p(r)] \frac{\sin(Q r)}{Qr} r^2 dr.
\label{eq:ff}
\eeq
From \eref{eq:ff} we see that since $(1-4\sstw)$ is small, a neutrino scattering elastically with a spin zero nucleus couples mostly to the neutron distribution.  A measurement of the cross section for this process provides a measurement of the neutron form factor.  Such a measurement would be complementary to the parity violating experiment with lead because it would provide additional data, obtained at different energy ranges and with different nuclei, that could be used to calibrate nuclear structure calculations.
 
To illustrate our idea, we consider the scenario of a nuclear recoil detector at a stopped pion neutrino source.  However, the idea is general and an experiment could be preformed at a low energy beta beam\cite{CV1,CV2} as well. (Using a nuclear recoil detector at a low energy beta beam was considered in References \cite{bueno,AM,BMR2}.) For our example scenario, we choose the CLEAN detector\cite{clean} and the decay at rest neutrino source at the Oak Ridge SNS\cite{sns}.
Such an experimental setup was proposed in \mycite{scholberg}.  Note that the idea of using neutrino-nucleus elastic scattering to measure the neutron form factor could have been conceived years ago, when the neutrino's existence and the neutral current were verified. However, the idea is practical in this age because neutrino beams and nuclear recoil detectors are becoming a reality.

The paper is organized as follows.  In Sec. II, we discuss the neutrino flux at the Oak Ridge facility as well as the expected capabilities of the CLEAN detector. In Sec. III, we present the general formalism for calculating number of events in a detector and present events in the detector predicted by different analytic models for the form factor. In Sec. IV we analyze the potential for a detector to distinguish between form factors from different nuclear structure calculations. 
We give conclusions in Sec. V.

\section{Neutrino Source and Detector}
The spallation neutron source (SNS) at Oak Ridge National Lab will produce neutrons by firing a pulsed proton beam at a liquid mercury target. Among the fragments emitted will be pions. Negative pions will capture in the target, but positive pions will come to rest and decay into positive muons and muon neutrinos. The muons will also come to rest and decay, emitting a positron, electron neutrino and muon anti neutrino. The pion decay time is $26{\rm ns}$, and the muon decay time $2.2 \mu{\rm s}$.  The neutrinos will be emitted isotropically from the target. The energy spectra of the $\nu_e$ and $\bar{\nu}_\mu$ are precisely known since these neutrinos come from decay at rest; the $\nu_\mu$ will be mono-energetic. The nu-SNS (neutrinos at the SNS) proposal\cite{sns} calls for putting a neutrino detector in a room about 20 meters from the mercury target. The direction of the proton beam to the target defines the forward direction; the neutrino detector would be located to the rear of the target, perpendicular to the beam line. In the rear location, the background of neutrinos coming from any pion or muon decay in flight in the forward direction would be negligible. The pulsed nature of the beam will be advantageous in reducing other types of background. The time difference for the emission of $\nu_e$'s  and $\bar{\nu}_\mu$'s
may also be used for distinguishing between events coming from these neutrinos and $\nu_\mu$'s as well as further reduction of background.  An expected flux of $10^7$ neutrinos per sec per 
${\rm cm}^2$ of each flavor is expected at a distance of 20 meters from the target, where the neutrino detector will be placed.
  
The CLEAN (cryogenic low energy astrophysics with noble gases) detector\cite{clean} was conceived for the purpose of detecting dark matter particles and low energy neutrinos. The detector will contain a liquid noble gas. When an elastic scattering event occurs in a liquid noble gas, ultraviolet scintillation light is given off. The CLEAN detector will use wavelength shifting film to convert ultraviolet light to visible light that can be detected by photomultipiers. The detector will be able to detect neutrino-electron and neutrino-nucleus elastic scattering events. 
Some neutrino applications include measuring the p-p solar neutrino flux, measuring neutrino cross sections to search for neutrino magnetic moment contributions, and detecting neutrinos from a supernova.  The detector concept is currently undergoing testing.  A full sized detector is expected to have a mass of 10-100 tons of liquid neon or argon.\footnote{See recent talk 
``SNOLAB workhop.pdf'' posted on \cite{snolab}.}   

An experiment to use a CLEAN detector at the SNS has recently been proposed\cite{scholberg}.
The neutrino-nucleus elastic scattering process has never been observed, and so a first goal of this experiment would be to detect this process. Detection of this process could also be used to search for beyond standard model physics by measuring \sstw, discovering or constraining non-standard neutrino interactions by looking for their contributions to the cross section, and looking for a neutrino magnetic moment contribution to the cross section. The potential of these physics applications was studied in \mycite{scholberg} and it was found that an experiment would
be most sensitive to search/constrain non-standard neutrino interactions.

\section{Formalism to Calculate Events}
We now study the potential of an experiment using a CLEAN detector at the SNS to constrain nuclear physics theories.
We demonstrate this by presenting calculated numbers of neutrino-nucleus elastic scattering events in a detector.
Different nuclear models predict different neutron distributions, and hence different form factors for the cross section.
Therefore, given the neutrino flux distribution, different nuclear models will predict different numbers of events to occur in a detector.  The questions we address are: how much do the event rate curves differ according to different nuclear structure calculation predictions for the form factor, and what sensitivities must the experiment achieve to distinguish between these theories?

We consider events of $\nu_e$ and $\bar{\nu}_\mu$ scatterings.  These neutrinos come from the muon decay at rest.  Their spectra can be calculated from the expression for the differential decay rate of the muon (see \mycite{griffiths}) by integrating over the energies of the electron and $\bar{\nu}_\mu$ for the $\nu_e$ spectra, and integrating over the energies of the electron and $\nu_e$ for the $\bar{\nu}_\mu$ spectra. The normalized spectra are 
\bea
f_{\nu_e} = \frac{96}{m_\mu^4}(m_\mu E_{\nu_e}^2 - E_{\nu_e}^3) dE_{\nu_e} \\
f_{\bar{\nu}_\mu} = \frac{16}{m_\mu^4}(3 m_\mu E_{\bar{\nu}_\mu}^2 - 4 E_{\bar{\nu}_\mu}^3) dE_{\bar{\nu}_\mu},
\eea
where $m_\mu$ is the muon mass.  These expressions give the probability that a neutrino is emitted with energy in the range $(E, E+dE)$. Plots showing the spectra shape can be found in \mycite{scholberg}.  The number of events are obtained by folding the neutrino flux with the differential cross section.  In particular, the number of events per nuclear recoil energy is 
\beq
\frac{dN}{dT}(T) = N_t \, \mathcal{C}\int^{m_\mu/2}_{E_{\rm min}(T)} f(E)\, \frac{d\sigma}{dT}(E,T)\ dE,
\eeq
where $E_{\rm min}(T) = \frac 1 2 (T + \sqrt{T^2 +2TM})$ is the minimum energy a neutrino must have in order to be able to give the nucleus a recoil energy $T$, $N_t$ is the number of target nuclei in the detector, and $\mathcal{C}$ is the total number of neutrinos per second per cm squared of a given flavor reaching the target.  Multiplying the spectra $f(E)$ by $\mathcal{C}$ gives the neutrino flux at the detector.

We do not use form factors from nuclear structure calculations.  Rather, we represent different density distributions using analytic expressions which come from \mycite{engel}.  The density distribution from \mycite{engel} models the nucleons as having a constant interior density and surface thickness $s$.  The Fourier transform of the density distribution gives the form factor\cite{engel}
\beq
F(Q^2) = \frac{3j_1(Q R_0)}{Q R_0}\exp[-\frac 1 2 (Q s)^2], \label{eq:analyticFF}
\eeq
where $R_0^2= R^2-5s^2$, $R$ is the radius of the nucleus, and $s$ is the surface thickness of the nucleus. The momentum transfer $Q^2$ has been given above.  We are treating the proton and neutron distributions separately, as in \eref{eq:ff}.
The form factor we use is
\beq
F(Q^2) = \frac 1 {Q_w} [N F_n(Q^2) - Z (1 - 4 \sstw) F_p(Q^2)] \label{eq:sepFF}
\eeq
were $F_n(Q^2)$ and $F_p(Q^2)$ are the neutron and proton from factors, respectively.  We use the functional form of \eref{eq:analyticFF} evaluated at the neutron RMS radius $R_n$ and the proton RMS radius $R_p$ for the neutron and proton form factors, respectively.  

We consider the case of a detector filled with argon isotope $^{40}{\rm Ar}$, with $Z=18$ and $N=22$. An experimentally determined value for the mean square charge radius of this isotope is $<R_p^2>\approx 11.75$ \cite{blaum}. Therefore, we take the RMS value of the proton radius to be $R_p = 3.43$. For each form factor we use the same value of $s=0.5\,{\rm fm}$ for the surface thickness of the density distribution.  Of course, if the experiment is to be performed, one would use a specific nuclear structure calculation to self consistently model the nucleus and all parameters required to fit the data. The purpose of our calculation is to demonstrate the potential of doing such an experiment. It is sufficient to describe the nucleus with an analytic model and assign values for the parameters as we have done.  (We have in fact modified the parameters slightly and found our results and conclusions do not change.)

We are interested in using a measurement of the neutron form factor to distinguish between different nuclear structure calculations that predict different neutron distributions.  We are modeling the neutron distribution with an analytic expression.  The parameter in this expression that represents different theories is the neutron RMS radius.
Thus we will predict what events are expected in the detector, depending on what nature actually has chosen for the neutron distribution in the $^{40}{\rm Ar}$ isotope, by varying the neutron radius.

A current figure of merit for the neutron radius is that it is known to within an uncertainty of $10 \%$ for medium to heavy nuclei, though there is debate in the literature on this issue\cite{HPSM}.  For a nucleus with equal numbers of protons and neutrons, one may expect that the neutron and proton density distributions are equal and that $R_n \approx R_p$.  However, because of Coulomb repulsion, the proton radius may be larger.  For a large nucleus, where there are more neutrons than protons, we expect that the neutron radius will be larger. For our calculations, we first set $R_n=R_p$ and then vary $R_p$ by $\pm 20 \%$, $\pm 15 \%$ and $\pm 10 \%$. Even though neutron radii are likely known to $10 \%$, it will be useful exercise for us to consider the cases of $R_n=R_p\pm 20 \%$ and $R_n=R_p\pm 15 \%$, as will be seen in the next section. 

\begin{figure}
\scalebox{1.2}{\includegraphics{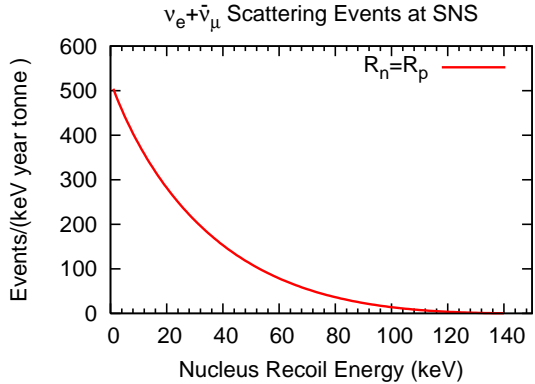}}
\caption{Number of $\nu_e+\bar{\nu}_\mu$ events in detector in units of number per (keV year tonne) assuming $1\times10^7$ neutrinos per second of each flavor are emitted from the source. Neutron radius is set equal to proton radius. } \label{fig:events}
\end{figure}

In \fref{fig:events} we show an example of events in the detector.  This plot shows the predicted total number of events that will occur from $\nu_e$ and $\bar{\nu}_\mu$ elastic scattering, per keV, per year, per tonne of detector material, plotted against nuclear recoil energy in keV. (Note that tonne = metric ton.) The numbers for this plot were calculated for the case $R_n=R_p$.  Events are plotted per tonne so that we may be general and not impose a size for the detector.

\begin{figure}
\scalebox{1.2}{\includegraphics{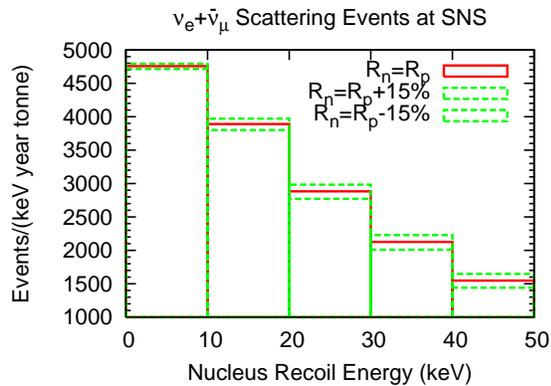}}
\caption{Number of  $\nu_e+\bar{\nu}_\mu$ events per (keV year tonne) summed in 10 keV bins. Note the range: only the first five bins are shown and the verticle axis starts at 1000 events. The solid line shows events predicted for the case $R_n = R_p$ and the dashed lines show predicted events for the cases $R_n = R_p \pm 15 \%$} \label{fig:binevents}
\end{figure}

Next we consider events for $R_n=R_p\pm 15 \%$ and compare these to events in the case $R_n=R_p$.  In \fref{fig:binevents}, events have been summed and placed in 10 keV bins; the first five bins are shown.  The solid line shows total events in each bin for the case $R_n=R_p$, and the dashed lines show total events in the cases $R_n=R_p\pm 15 \%$.  The difference in binned event rates for these cases is visible in the plot.  We can see that for low momentum transfers, i.e. low recoil energy, the events converge.  This is consistent with the normalization condition on the form factor, $F(Q^2=0)=1$.  (The matrix element of the vector current for a nuclear state, in the zero momentum transfer limit, is equal to the conserved charge associated with that current\cite{afg}. Since we have factored out the charges, see for example, \eref{eq:sepFF} and \eref{eq:cross-section}, the proper condition for our expression in this limit is normalization to 1.) The low energy threshold of the detector is expected to be about 10 keV, i.e. the detector is not expected to be able to see nuclear recoil events with energies below 10 keV.  When analyzing the potential of this experiment to distinguish nuclear structure calculations for the from factor, we will consider events above the 10 keV threshold.  However, later we will comment on opportunities for distinguishing theories if the threshold of the detector can be made lower.

As there are thousands of events per bin in \fref{fig:binevents} the scale makes it difficult to see how the events differ over the whole range of nuclear recoil energy. Therefore, we present Table \ref{tab:binevents} which shows for each bin range the number of predicted events for each of the cases $R_n = R_p$, $R_n = R_p\pm 20\%$, and $R_n = R_p\pm 10\%$.  The table also shows the percentage difference in events for the modified cases of $R_n$ compared to the case $R_n = R_p$.  

\section{Analysis}
Using the table we can best gauge the potential for the experiment to be used to fit nuclear structure calculations for the form factor.  The fit will depend on what the final error bars are for the data.  Guided by the discussion in \mycite{scholberg}, we consider the scenario of $10 \%$ systematic uncertainty on the data.  This can take into account uncertainties in the proton beam, neutrino flux reaching the detector, and the detector.  Note that this systematic uncertainty need not apply to all energy bins.  Events in some bins may be resolved better or worse. In our analysis we will consider some effects of backgrounds and detector efficiency.  We remark here that backgrounds for some bins may be more or less severe, and the detector might have better or worse efficiency for certain energies.  It should be kept in mind that the true potential of measuring the neutron form factor with this type of experiment can be evaluated once backgrounds, detector efficiencies, and systematic uncertainties are better known. 

We now consider the case of $R_n=R_p\pm 20\%$.  From the table, we see that events in the bins with recoil energy $> 70\,{\rm keV}$ differ from events for the case $R_n=R_p$ by more than 
$10 \%$.  Therefore, if the systematic uncertainty on the data in these bins is $10\%$, and there are enough events for low statistical uncertainty, then there would be potential for the data to be fit to nuclear models that predict the neutron radius to have values $R_n=R_p\pm 20\%$.  Even if background cuts and detector efficiency would remove some of the events from being used, it is likely that there is still a large enough difference between events from the models that they could be distinguished.  For example, in the $80-90\,{\rm keV}$ bin, there is a difference of about 50 events between the cases $R_n=R_p\pm 20\%$ and $R_n=R_p$.  This should leave enough room for some adjustments for background cuts and efficiency.  

The exercise of considering the case $R_n=R_p\pm 20\%$ is useful because it demonstrates that this type of experiment could be used to measure a neutron distribution.  However, we considered a specific experimental setup for our example, and the capabilities and precision of this setup are still being determined.  Obviously, if the experiment is more sensitive, a better measurement of the neutron distribution can be made.  Therefore, we next consider what precision would be required to determine $R_n$ to better than $10\%$, the current figure of merit for how well the neutron radius is known.

Consulting the table, we see that for the case $R_n=R_p\pm10 \%$, the events in the bins with recoil energies $> 100\,{\rm keV}$ differ from the events for the case $R_n=R_p$ by more than $10 \%$.  Therefore, if the systematic uncertainty on the data in these bins is $10 \%$, then one might think the data can be used to distinguish nuclear structure calculations that predict the neutron radius to have values $R_n=R_p\pm10 \%$, from theories that predict $R_n\approx R_p$.  However, in these energy  bins, the models $R_n=R_p\pm10 \%$ differ from the model $R_n\approx R_p$ by only a few events.  Since the detector will not be $100 \%$ efficient, and background events will need to be accounted for, our calculation indicates these bins will not provide useful information for fitting to theory.  We therefore suggest a goal for the systematic uncertainty needed to measure the neutron form factor of $^{40}{\rm Ar}$ with enough precision to extract $R_n$ to an accuracy better than $10 \% $. With our example for an experimental setup and assuming enough events to have a low statistical uncertainty, the systematic uncertainty would have to be reduced to better than $5\%$.  If this were the case, then data from bins $> 50\,{\rm keV}$ could be used.  For example, our calculation indicates a difference of about $\pm 50$ events between the cases $R_n=R_p\pm10 \%$ and $R_n=R_p$ in the bin $60-70\,{\rm keV}$, with percentage difference of $> 6\% $.  This suggests that if systematic uncertainties can be reduced to $5\%$ or better, then even when background and detector efficiency is accounted for, the data can still be used extract $R_n$ to better than 
$10 \%$.

Note that the calculated events we presented are given per year, per tonne of detector material. If data is taken longer, or the detector mass is larger than 1 tonne, more events will be available and hence statistical uncertainies will be reduced.  Note also that other nuclei, for example Ne or Xe, could be used in the detector.  The prospects of doing a neutron form factor measurement with other nuclei should be investigated.  We remark again that our analysis was done for a particular example of an experimental setup.  A different experimental configuration will obviously have a different potential for doing this measurement.

In our analysis we have ignored contributions to the cross section from beyond standard model physics.  However, it is possible that new physics exists in nature.  If this is the case then how could we know if a difference in measured events from predicted events stems from a modification to constants in the cross section or a modification to the nuclear form factor?  One feature at our disposal is that different form factors will have slightly different shapes, while a correction to say, \sstw, would shift the cross section.  A shift in the cross section would shift the number of events at all recoil energies, including the lowest recoil energy.  However, a correction to the form factor would not effect the low recoil energy events. (Recall the form factor is normalized to one at zero momentum transfer.)  Such differences in the event rates could be observable if the threshold of the detector can be made lower.

\section{Conclusions}
We have considered a new idea for a method to measure the neutron form factor of a nucleus --- by detecting neutrino-nucleus elastic scattering.  We illustrated this idea using a particular example of an experimental configuration --- the neutrino source at the Oak Ridge SNS and a CLEAN detector filled with one tonne of liquid argon.  We found that for this setup, it is possible to measure the neutron radius of the argon isotope $^{40}{\rm Ar}$ to better than $10 \%$ if systematic uncertainties could be reduced to better than $5 \%$.  The idea is general in the sense that other nuclei could be used in the detector, or a different low energy neutrino source could be used. 

\section{Acknowledgments}
The authors thank J.~Carlson, J.~Engel, S.~Reddy, G.~Rupak, K.~Scholberg, T.~Schaefer and A.~Young for useful discussions.
This work was supported by the U.S. Department of Energy under Grant No. DE-FG02-02ER41216.

\begin{table*}
\begin{tabular}{c|c|c|c|c|c|c|c|c|c|c}
Bin Range & $R_n = R_p$ & $R_n = R_p$ & $\%$ diff. & $R_n = R_p$ & $\%$ diff. &
	$R_n = R_p$ & $\%$ diff. & $R_n = R_p $ & $\%$ diff. \\
(keV) & & $+ 20\%$ & & $ - 20\%$ & &  $ + 10\%$ & & $ - 10\%$ & \\
\hline
   0-10 & 4756 & 4714 &   -1 & 4793 &    1 & 4729 &   -1 & 4781 &    1 \\
  10-20 & 3891 & 3800 &   -2 & 3971 &    2 & 3832 &   -2 & 3946 &    1 \\
  20-30 & 2884 & 2772 &   -4 & 2984 &    3 & 2811 &   -3 & 2952 &    2 \\
  30-40 & 2126 & 2010 &   -5 & 2230 &    5 & 2050 &   -4 & 2196 &    3 \\
  40-50 & 1549 & 1441 &   -7 & 1647 &    6 & 1478 &   -5 & 1616 &    4 \\
  50-60 & 1110 & 1015 &   -9 & 1197 &    8 & 1048 &   -6 & 1169 &    5 \\
  60-70 &  778 &  700 &  -10 &  851 &    9 &  726 &   -7 &  827 &    6 \\
  70-80 &  529 &  468 &  -12 &  587 &   11 &  489 &   -8 &  568 &    7 \\
  80-90 &  347 &  301 &  -13 &  390 &   13 &  316 &   -9 &  376 &    8 \\
  90-100 &  215 &  184 &  -15 &  246 &   14 &  194 &  -10 &  236 &   10 \\
 100-110 &  124 &  104 &  -16 &  144 &   16 &  111 &  -11 &  137 &   11 \\
 110-120 &   64 &   52 &  -18 &   75 &   18 &   56 &  -12 &   71 &   12 \\
 120-130 &   27 &   22 &  -19 &   32 &   19 &   23 &  -13 &   30 &   13 \\
 130-140 &    8 &    6 &  -21 &    9 &   21 &    6 &  -14 &    9 &   14 \\
\end{tabular}
\caption{Table shows binned events. All events are given in units of per (keV tonne year). Bins ranges are given in the first column in keV. The second column shows the number of events in each bin predicted by our model for the case $R_n=R_p$. The third column shows the predicted number of events for the case $R_n=R_p+20\%$. The fourth column shows the percentage difference between events for the case $R_n=R_p+20\%$ and the case $R_n=R_p$. The fifth column shows the predicted number of events for the case $R_n=R_p-20\%$; the sixth column shows percentage difference between events for this case and $R_n=R_p$. The eighth and tenth column show events for the cases $R_n=R_p \pm 10 \%$, while the ninth and eleventh columns show the percentage difference for the events of these cases and the case $R_n=R_p$. All numbers are rounded to the nearest whole number.}
\label{tab:binevents}
\end{table*}


\newpage
\begin{thebibliography}{11}
\bibitem{HPSM}C.~J.~Horowitz, S.~J.~Pollock, P.~A.~Souder, and R.~Michaels, Phys. Rev. C {\bf 63}, 025501 (2001).
\bibitem{furnstahl}R.~J.~Furnstahl Nucl. Phys. {\bf A706}, 85 (2002).
\bibitem{prex}http://hallaweb.jlab.org/parity/
\bibitem{freedman}D.~Z.~Freedman, Phys. Rev. D {\bf 9}, 1389 (1974).
\bibitem{FST}D.~Z.~Freedman, D.~N.~Schramm, and D.~L.~Tubbs, Ann. Rev. Nucl. Sci. {\bf 27}, 167 (1977).
\bibitem{BMR}J.~Barranco, O.~G.~Miranda, and T.~I.~Rashba, JHEP {\bf 12}, 21 (2005).
\bibitem{HCM}C.~J.~Horowitz, K.~J.~Coakley, and D.~N.~McKinsey, Phys. Rev. D {\bf 68}, 023005 (2003).
\bibitem{CV1}C.~Volpe, J. Phys. G {\bf 30}, L1 (2004).
\bibitem{CV2}C.~Volpe, J. Phys. G {\bf 34}, R1 (2007).
\bibitem{bueno}A.~Bueno, M.~C.~Carmona, J.~Lozano, and S.~Navas, Phys. Rev. D {\bf 74}, 033010 (2006).
\bibitem{AM}P.~S.~Amanik and G.~C.~McLaughlin, Phys. Rev. D {\bf 75}, 065502 (2007).
\bibitem{BMR2}J.~Barranco, O.~G.~Miranda, and T.~I.~Rashba, arXiv:hep-ph/0702175.
\bibitem{clean}D.~McKinsey and K.~J.~Coakley, Astropart. Phys. {\bf 22}, 355 (2005).
\bibitem{sns}http://www.phy.ornl.gov/nusns
\bibitem{scholberg}K.~Scholberg, Phys. Rev. D {\bf 73}, 033005 (2006).
\bibitem{snolab}http://mckinseygroup.physics.yale.edu/publications/
\bibitem{griffiths}D.~J.~Griffiths, Introduction to Elementary Particles, John Wiley and Sons Ltd., New York, 1987.
\bibitem{engel}J.~Engel, Phys. Lett. B {264}, 114 (1991).
\bibitem{blaum}K.~Blaum, W.~Geithner, J.~Lassen, P.~Lievens, K.~Marinova, and R.~Neugart, Hyperfine Interactions
\bibitem{afg}P.~S.~Amanik, G.~M.~Fuller, and B.~Grinstein, Astroparticle Physics {\bf 24}, 160 (2005).
{\bf 162}, 101 (2005).
\end{thebibliography}
\end{document}